# A magnetotelluric image of the Curnamona Province and the adjacent Delamerian Orogen margin – new insights into the crustal architecture

Author #1: Wenping Jiang, Geoscience Australia, Canberra, Australian Capital Territory, Australia, Wenping.Jiang@ga.gov.au

Author #2: Michael Doublier, Geoscience Australia, Canberra, Australian Capital Territory, Australia, Michael.Doublier@ga.gov.au

Author #3: Russell Korsch, Geoscience Australia, Canberra, Australian Capital Territory, Australia, russell.korsch@ga.gov.au

Author #4: Andy Clark, Geoscience Australia, Canberra, Australian Capital Territory, Australia, Andrew.Clark@ga.gov.au

Author #5: Malcolm Nicoll, Geoscience Australia, Canberra, Australian Capital Territory, Australia, malcolm.nicoll@ga.gov.au

Author #6: Adrian Hitchman, Geoscience Australia, Canberra, Australian Capital Territory, Australia, adrian.hitchman@iinet.net.au

Author #7: Yanbo Cheng, Geoscience Australia, Canberra, Australian Capital Territory, Australia, yanbo.cheng@ga.gov.au

Note: this paper is currently under consideration of Earth, Planets and Space

**Abstract**

We have used new magnetotelluric data collected in the Curnamona Province and the adjacent part of the Delamerian Orogen margin to image electrical conductivity structures and to inform the understanding of the crustal architecture within the regional geological context. Deep-crustal seismic reflection data has been integrated to better image the positioning of major structures at the first-order crustal scale. Resistivity models were derived from inverting 231 magnetotelluric sites on an approximately 20-25 km grid supplemented by denser sites across known geological structures and along reflection seismic lines. The preferred 3D resistivity model confirms, and resolves in greater detail, crustal-scale conductive features that have been mapped by the long-period data collected at half-degree spacing as part of the Australian Lithospheric Architecture Magnetotelluric Project (AusLAMP), that is, the prominent Curnamona Province Conductor and the two Nackara Arc conductors. The new model reveals that the eastern Nackara Arc (ENAC) conductor continues as the Broken Hill Conductor (BHC) into the Curnamona Province. Regional geological considerations suggest that its formation is possibly linked to rifting/extension in the early Cambrian. Although we recognise that the east-west trending Wilcannia Conductor could be a possible continuation of the ENAC-BHC zone, integration with recently acquired deep seismic reflection data and evaluation of the geological setting lead us to suggest that they are not genetically linked. We suggest that the Wilcannia Conductor is younger and most likely is related to late Delamerian (~500 Ma) or Siluro-Devonian magmatism. Finally, these conductivity anomalies may represent large-scale trans-crustal structures that control the emplacement of low volume alkaline ultramafic magmas, and show a spatial relationship with certain mineral deposit types, suggesting a possible control on the distribution and formation of metallogenic provinces/belts in the region. This will be further investigated in future work.

# 1 Introduction

Magnetotellurics (MT) is recognised as a powerful geophysical method for geological interpretation and mineral prospectivity assessment on multiple scales (e.g., Heinson et al. 2022, Jiang et al. 2022, Kay et al. 2022) where electrical conductivity/resistivity heterogeneity of the subsurface places constraints on lithospheric architecture and possible mineralogical compositions. Recent studies have demonstrated remarkable correlations between crustal and mantle conductive regions and known mineral deposits, e.g., orogenic gold deposits in southeastern Australia (Heinson et al. 2021) and southeastern Tibet (Hou et al. 2022), and Iron Oxide Copper-Gold (IOCG) deposits along the eastern margin of Gawler Craton in South Australia (Heinson et al. 2006, Heinson et al. 2018, Skirrow et al. 2018) and along the eastern margin of the Mount Isa Province in Northern Australia (Wang et al. 2018, Jiang et al. 2019). Kirkby et al. (2022) have used datasets from Australia, North and South America, and China to quantitatively assess spatial correlations between conductive regions and mineral deposits formed in convergent margin settings. They found that deposits formed at later stages of an orogenic cycle, e.g., orogenic gold, show strong correlations with crustal conductors, whereas volcanic hosted massive sulphide (VHMS) and porphyry copper deposits that formed at early stages of an orogenic cycle show weak to moderate correlations with conductors in the upper mantle. This empirical assessment needs to be further tested with new data and observations, for example, more recently, new evidence reveals clear spatial correlation between porphyry copper/copper deposits and crustal-scale conductivity anomalies (Sheng et al. 2024, Kay et al. 2025).

The ~830 to ~495 Ma Delamerian Orogen, the oldest part of the Tasmanides of eastern Australia (Glen 2005), wraps around the Proterozoic Curnamona Province in New South Wales (NSW) and South Australia (SA), and extends through western Victoria to western Tasmania. The regional scale Tasmanides is illustrated in Figure 1 by Glen (2013). The Delamerian Orogen represents initially a Neoproterozoic rifted continental margin that converted to a convergent margin setting in the Cambrian, with west-dipping subduction and arc magmatism (Cayley et al. 2018, Foden et al. 2020), marking the transition from Proterozoic Australia to the Phanerozoic Tasmanides. Subduction was terminated by the Delamerian Orogeny in the mid-to-late Cambrian, with the locus of tectonic activity subsequently migrating to the east into the Lachlan Orogen (Glen 2005, Glen 2013). The orogen was also affected by younger both contractional and extensional tectonic events related to younger orogenic cycles effecting the Tasmanides (Gilmore et al. 2023, Glen 2013). Therefore, the orogen and its mineral systems present a complex amalgam of overlapping passive margin, convergent margin, intraplate and distal backarc systems that span the Neoproterozoic to Devonian. The orogen is prospective for an array of minerals, including gold, copper and critical minerals (Gilmore 2010, Huston et al. 2016, Duncan et al. 2018, Ford et al. 2018, Gilmore et al. 2023, Cheng et al. 2024). To the west, the exposed Paleo-Mesoproterozoic Curnamona Province is separated from the Archean-Proterozoic Gawler Craton by the Adelaide Rift Complex (Preiss 2000, Counts 2017). The Province hosts significant Broken Hill-type lead-zinc-silver- deposits and has high prospectivity for Olympic Dam-style iron-oxide copper gold uranium (IOCG-U) mineralisation (Conor and Preiss 2008).

Previously, crustal-scale conductivity anomalies were imaged in the centre of the Curnamona Province, and in the Nackara Arc (named after its arcuate shape) region to the south, using long-period MT data collected at a ~55 km grid spacing, as part of the

Australian Lithospheric Architecture Magnetotelluric Project (AusLAMP) (Robertson et al. 2016, Robertson et al. 2020). Follow up surveys were undertaken to characterise the Curnamona Conductor, including a 2 km spacing broadband transect that intersected the conductor (Kay et al. 2022). Most recently, two Broadband MT surveys were undertaken by the University of Adelaide/AuScope and Geoscience Australia, the Curnamona Cube and the Curnamona Cube Extension surveys (Figure 1). Both surveys have been acquired with a grid spacing of ~20-25 km and complement the existing AusLAMP dataset by imaging the electrical resistivity structures of the area in greater detail, including in the uppermost crust, allowing for better utilisation of surface geological constraints for interpretation, and investigation of aspects of the existing AusLAMP models where interpretation was less certain. Here we use the data of these two surveys to image resistivity structures of the Curnamona Province and the adjacent part of the Delamerian Orogen margin; and to combine interpretations of existing AusLAMP models to investigate the regional crustal and lithospheric architecture, and to provide new timing constraints on the formation of major regional conductors. Recently acquired deep-crustal seismic reflection data along line 22GA-CD2 (Doublier et al. 2024) has been integrated with the MT data to better understand the positioning of regional conductors within geological context at the first-order crustal scale.

## 2 Survey and Data

The Curnamona Cube survey was undertaken by the University of Adelaide/AuScope in 2022. Broadband MT data in the range of 0.005-1000 s were acquired at 132 sites on an approximately 25 km grid with vertical magnetic data collected at 28 sites. For details of data acquisition and processing, see the upcoming data release and associated publications (Heinson et al. 2024, Kay et al. 2025).

Geoscience Australia undertook data acquisition and processing for the Curnamona Cube Extension survey in early 2023. Broadband MT data were acquired at 99 sites on an approximately 20 km grid, supplemented by densely spaced (~10 km) sites across known geological structures and along reflection seismic lines (Geoscience Australia 2023). Instruments recorded five channels (three magnetic and two electric fields), for a minimum of 24 hours, with a target bandwidth of 0.0001–1000 s. Processed data show good quality at the majority of the survey sites (Figure 2), except for a small number of sites affected by environmental and cultural noise. Details of data acquisition and processing are reported by Jiang et al. (2023).

We derived the preliminary geoelectric structure from the data using phase tensor (Caldwell et al. 2004) and induction vector (Parkinson 1959) analyses. The phase tensor, defined as the ratio of the real and imaginary parts of the impedance tensor, reflects the underlying regional geoelectric structure (Booker 2014). Phase tensor maps at 1 s and 500 s are shown in Figure 3 where phase tensor ellipses are coloured by the minimum phase, indicating how the resistivity changes with depth. When the minimum phase is greater than 45°, for example at 500 s (Figure 3b), in the centre of the Curnamona Province, the subsurface is becoming more conductive with increasing depth within the lower crust. In the southeast part of the survey area the minimum phase at some sites exceeds 45° indicating a transition to a more conductive lower crust in the area. Induction vectors represent changes in the vertical magnetic field due to lateral variations in electrical conductivity. By Parkinson convention, real induction vectors point towards zones of higher conductivity and away from resistive zones. At short period of 1 s, induction vectors are small, reflecting low intensity anomalous current concentration at shallow depth. At 500 s, induction vectors are greater, indicating a higher intensity or magnitude of conductivity gradient at depth. The induction arrows in the southeast of Delamerian Orogen point toward the centre of the Curnamona

Province, and those in the southwest point toward the Nackara Arc in the southwest (shown in Figure 1).

## 3 Inversion and Resistivity Model

We inverted the MT data acquired at 231 sites by the Curnamona Cube and Curnamona Cube Extension surveys. We jointly inverted the full impedance tensor and tipper where data was available. We used the 3D ModEM code (Egbert and Kelbert 2012, Kelbert et al. 2014) and applied the standard minimum-structure non-linear conjugate gradient algorithm to solve the regularised inversion problem.

Multiple modelling tests were performed to determine the optimal model parameters and to ensure the robustness of major features discussed below. These include mesh resolution and rotation, starting model, data subset, smoothing factor/covariance, and error floor. A summary of the results is given in the Supplement section "Model Parameterisation and Sensitivity Tests".

To improve 3D modelling efficiency, the array of sites was rotated 30° anticlockwise to align with the model grid, and the data were rotated clockwise to keep the same geological strike angle with reference to the source field polarisation directions. The preferred model consisted horizontally of 222 × 168, 3 km sized cells and 9 padding cells extending in all directions. The size of the padding cells increased by a factor of 1.4 with distance from the model edge. Vertically, there were a total of 90 layers. The first layer thickness was set as 20 m, and subsequent layer thicknesses were increased by a factor of 1.2 to a depth of about 800 km. Data including impedance tensors and tippers at 37 periods in the range of 0.001–1000 s were inverted, where data was available. Noisy data with large standard deviations was excluded. Error floors of 5% of the square root of $Z_{xy}Z_{yx}$ (where $Z_{xy}$ and $Z_{yx}$ are the two off-diagonal impedance tensor components) were applied to each impedance tensor component, and an absolute value of 0.02 was applied to the tipper. The starting model was a uniform 100 Ω·m half space. A covariance value of 0.3 was applied twice in all directions across the model cells and was increased to 0.5 after 79 iterations to further smooth the model. The preferred model converged to a noise-normalised root-mean-square (RMS) misfit value of 2.14 after 142 iterations, which we considered to be a reasonably good fit to the data across most of the area (Figure 4). It is often challenging to fit data well in a geologically complex area, for example, the southern boundary of Curnamona Province, where MT responses are also complex. In part, it is due to the larger site spacing and lower model resolution, compared to the variation of resistivity structures. Supplement Figure S3 gives comparisons of modelled and observed data at nine representative sites.

The preferred resistivity model is presented in Figure 5**Error! Reference source not found.**, including depth slices at ~10 km, ~20 km, ~30 km, ~40 km and vertical slices within model cartesian space. The prominent conductive zones observed in the centre of Curnamona Province, and the Nackara Arc in the southwest are consistent with data analysis (Figure 3b).

Although this study focuses on the broadband MT survey and interpretation, we have inverted long-period data at 110 AusLAMP sites that are available to this study (locations shown in Figure 1), together with the broadband data to better constrain the crustal and possibly lithospheric scale structures in the region. The inversion model result is presented in Supplement section "Combined Long-period and Broadband Model".

## 4 Results and Interpretation

The large, west dipping conductor in the centre of the Curnamona Province (Curnamona Conductor) was imaged previously using long-period AusLAMP data (Robertson et al. 2016) and a high-resolution broadband MT transect (Kay et al. 2022).The reduced resistivity was interpreted as an enrichment of carbon by thermal and structural reworking of the province, in particular, due to subduction related refertilisation (Robertson et al. 2016, Brand et al. 2017), although, the timing is enigmatic. The AusLAMP model also imaged the Eastern Nackara Arc Conductor (ENAC) and the Western Nackara Arc Conductor (WNAC) at 30 - 80 km (refer to Fig. 5 and Fig. 6 in Robertson et al. (2016)). Robertson et al. (2016) suggested that the lithospheric scale Nackara Arc conductors may preserve a signature of mantle-sourced fluids or partial melt, which are likely to have intruded along pre-existing major faults or zones of weakness in the lithosphere. The new resistivity model presented here further confirms, and better resolves, crustal-scale conductive features in the region, shown in Figure 5, Figure S1, and model slices and SGid in Jiang et al. (2023). The Curnamona Conductor is characterised as a pseudo-circular conductor similar to the outline of Curnamona Province except the south and southeast parts. The conductor extends from a depth of ~40-50 km to ~3 km with some small parts connecting to the near surface. Please note, the base of the conductors could not be modelled precisely due to the diffusive nature of the electro-magnetic technique. The structure of this conductor lead us to agree with Kay et al. (2022) that this conductivity anomaly might represent a graphitic suture zone of Paleoproterozoic age that is similar to other conductors in Australia, for example, the Carpentaria Conductor along the eastern margin of the Mount Isa Province in northern Australia (Lilley et al. 2003, Wang et al. 2018, Jiang et al. 2019), and the Stuart Shelf Conductor along the eastern margin of the Gawler Craton in southern Australia (Heinson et al. 2018, Heinson et al. 2022). Elsewhere, crustal-scale conductivity anomalies have been mapped, e.g., the strike-slip Fraser Fault in British Columbia, Canada (Jones et al. 1992); the North American Central Plains conductivity anomaly (Camfield and Gough 1977, Jones et al. 1997, Jones et al. 2005); and enhanced conductivity has been attributed to deformation or mineralisation of graphitic or sulfidic sedimentary rocks associated with collisional processes during orogenesis (Boerner et al. 1996, Jones et al. 1997, Ferguson et al. 1999, Meqbel et al.

2014). Sulphides are considered unlikely forming significant crustal conductors because of their insufficient volume in Earth composition (Allègre et al. 1995). Therefore, they may only contribute to discrete, localised features in the upper crust. An interpretation of causality between enhanced conductivity and aqueous fluids and/or partial melt is only applicable to tectonically active regions, e.g., Tibetan Plateau in China (Hou et al. 2022, Sheng et al. 2024).

Our model also images the northeastern parts of the two Nackara Arc Conductors and reveals a shallower top (~5-10 km) and base (~50-60 km), compared to the AusLAMP model. Both the ENAC and WNAC do not continue into the Curnamona Province in the upper and middle crust, shown in Figure 5a, 5b and Figure S1a and S1b, and model slices in Jiang et al. (2023). Also, the WNAC does not have a well-defined continuation with the Curnamona Conductor in the lower crust and upper mantle (Figure 5d and Figure S1d). These two separate conductors may represent ancient fluid pathways related to the rifting cycles where the lithosphere was mechanically weakened and refertilised during metasomatism (Robertson et al. 2016). However, these pathways may not extend into the lower mantle, but only into the upper mantle.

The new model also reveals a continuous arcuate conductor at lower crust extending from the ENAC to Broken Hill. We name the northeastern continuation of the ENAC as the Broken Hill Conductor (BHC) (Figure 5c and 5d, and Figure S1c and S1d). Integration with the Delamerian AusLAMP model (Robertson et al. 2020) reveals that the ENAC and BHC together define a (semi-)coherent conductive zone that continues for > 500 km, with a consistent northeast to east-northeast trend. The northeastern part of this conductive zone runs sub-parallel to the regional structural grain of the Delamerian Orogen and major structures such as the Anabama Shear Zone and its extension into NSW, the Redan Shear Zone (refer to Figure 8 in Preiss (2000), Preiss (2019)). Nevertheless, the southern part, south of the apex of the Nackara Arc, shows a moderate angular relationship with the regional structural grain and the zonation of the Delamerian Orogen (Figure 6), which we consider, indicates that the development of the ENAC-BHC conductive zone is not linked to the arc formation and/or orogenesis. Within the regional geological context, the orientation of the conductive zone is most consistent with formation during northwest-southeast oriented rifting and/or extension, which occurred in the early Cambrian during the formation of the Kanmantoo Trough (Preiss 2000), and again in the Jurassic related to the final breakup between Australia and Antarctica (e.g., Teasdale et al. (2003)). This architecture is superimposed onto the Neoproterozoic strata and underlying extended Gawler basement (Wise et al. 2025), and has been used repeatedly as a pathway for low volume alkaline ultramafic magmas. The respective rocks are, in plan view, mostly located along the edges of the conductive zone, and include Ordovician lamprophyres (Jaques 2008, Bryant et al. 2023) and Jurassic kimberlites (Robertson et al. 2016, Sudholz et al. 2022). We interpret the lamprophyres to provide a minimum age constraint for the formation of the ENAC-BHC zone, which would be consistent with a genetic link to the early Cambrian rifting. However, a possible relationship to older, pre-existing structures related to the Proterozoic development of the mid- to lower crustal cratonic basement cannot be ruled out.

The Wilcannia Conductor has been interpreted previously in the AusLAMP model as a possible continuation of the ENAC-BHC zone (Kirkby et al. 2020). In our model, both conductors occur at similar crustal levels and are possibly connected (Figure 5 and Figure S1). Nevertheless, the Wilcannia Conductor shows several characteristics which need to be considered when assessing the relationship between both conductors. Most noticeable in plan view are the east-west orientation (sub-perpendicular to the trend of the orogen), and its position within the orogen extending from the arc into the forearc (Figure 6). Integration with the recent deep seismic reflection line 22GA-CD2 (total length of 178 km) provides further constraints on the geological setting and structural context (Doublier et al. 2024): Line 22GA-CD2 is located east of the Curnamona Province, representing a transect through the eastern part of the Delamerian Orogen (Figure 6). Under thin cover of the Cenozoic Murray Basin, the line images components of the Devonian Darling Basin system (Tararra-Menindee Trough, Blantyre Sub-basin, Poopelloe Lake Trough), Cambro-Ordovician sediments deposited after the Delamerian Orogeny (Mutawintji Group), Cambrian strata predating the Delamerian Orogeny (Cambrian arc package, Teltawongee and Ponto groups), and the Barrier Seismic Province that forms the lower and parts of the middle crust along the line, and is interpreted to consist of interbedded and/or structurally intercalated mafic igneous rocks and sedimentary rocks originally deposited in an accretionary wedge setting and adjacent forearc (Figure 7; Doublier et al. (2024)).

The crustal architecture along line 22GA-CD2 preserves many characteristics of the early evolution of westward Cambrian subduction, accretion and orogeny as part of the Delamerian Orogen, which is reflected in the coherent, south-westward dipping orientation of major structures (Figure 7b). The seismic line intersects the orogenic architecture at a moderate angle, resulting in relatively shallow apparent dip magnitudes. This initial configuration has been reworked and reactivated during later orogenic events, which result in the present-day compartmentalisation of the Darling Basin and the formation of intervening basement highs (Lake Wintlow Belt and Wilcannia High).

In the cross section along seismic line 22GA-CD2, low resistivity in the upper crust is related to the Darling Basin, and the Allambie Woolshed Granite in the Wilcannia High. The Wilcannia Conductor is located in the middle to lower crust and confined to the Barrier Seismic Province (Figure 7c). It appears not to be affected by the major crustal structures related to Delamerian subduction and orogeny during the middle to late Cambrian, suggesting a formation after at least most of this mid- to lower crustal architecture was established.

Considering these characteristics, there are some important differences between the Wilcannia Conductor and the ENAC-BHC zone: (i) the different orientation of the Wilcannia Conductor is difficult to reconcile with the proposed early Cambrian rift-related origin, and consistent trend, of the ENAC architecture; (ii) the ENAC-BHC zone is confined to cratonic basement of the Curnamona Craton and the Gawler Craton underlying Neoproterozoic and Cambrian strata, whereas the Wilcannia Conductor is located in very different (younger) crust of the Barrier Seismic Province; (iii) the early Cambrian rifting predates the structural architecture and at least some of the Barrier Seismic Province lithologies of accretionary affinity, that are, in turn, overprinted by the Wilcannia Conductor (Figure 7). Based on these differences, we suggest that a genetic relationship between the Wilcannia Conductor and the ENAC-BHC zone is unlikely, and that the Wilcannia Conductor is a younger feature.

The Wilcannia Conductor is broadly positioned in an area known as the Grasmere knee zone, where there is a ~90-degree change in the trend of the Delamerian Orogen, from northwest-southeast trends in the Koonenberry Belt to the north to northeast-southwest trends in the Loch Lilly-Kars Belt to the south. This change in trend has been attributed to oroclinal folding around the edge of the Curnamona Province in the later/final stages of the Delamerian Orogeny (Greenfield et al. 2011), although a substantial part of the curvature may likely be attributed to the initial rift margin geometry (Clark et al. 2024). It is feasible that a cryptic east-west trending structural corridor has been established during this event, although this is not clearly evident in line 22GA-CD2. In such a model, the Wilcannia Conductor could be related to magmatism and fluid flow during the formation of this corridor, and magmatism at ~500 Ma is present in the area (Mole et al. 2024). Alternatively, it could be related to younger Siluro-Devonian magmatism, which includes the nearby ~420 Ma Allambie Woolshed Granite (Black 2006).

A younger evolution cannot be ruled out, but seems less likely based on current data. The east-west orientation of the Wilcannia Conductor could be related to extension during the Devonian Darling Basin sedimentation (Cayley 2015), or basin inversion during the Kanimblan-Alice Springs Orogeny (~360 Ma to 340 Ma; Glen 2005; Doublier et al. 2024). Significant vertical movement during the latter, with offsets > 3.5 km is apparent in the line 22GA-CD2 (Figure 7), and contributed to the final establishment of the Wilcannia High, an east-northeast to west-southwest oriented basement high that is co-located with the central part of the Wilcannia Conductor. Nevertheless, there are no magmatic and/or fluid flow events recognised during that time as a potential explanation of the conductivity anomaly, and in 22GA-CD2 the younger structural events predominantly affect the upper crust, above the mid- to lower crustal WC, hence not providing further timing constraints.

At a more regional scale in the upper-mid crust, the resistivity model at ~10 km depth indicates a subtle, largely continuous conductive zone along the Delamerian Orogen margin (Figure 8). This provides geophysical evidence to support the geological interpretation that the Cambrian volcanic arc is likely to be a single, east-facing convergent margin developed in what is now the Delamerian Orogen of the Australian mainland (Greenfield et al. 2011, Cayley et al. 2018, Foden et al. 2020, Wise 2023, Clark et al. 2024). In addition, drill core analysis has revealed a broad footprint of ~515−500 Ma volcanic and intrusive igneous rocks of varying composition between the exposed Koonenberry Belt, the subsurface Loch Lilly-Kars Belt and the exposed Grampians-Stavely Zone further to the south in western Victoria (Mole et al. 2023, Mole et al. 2024). Our model therefore helps to delineate the search space for a range of mineral deposit types linked to Delamerian magmatic arc activity, such as porphyry-epithermal, volcanic-hosted massive sulfide and other deposit types of magmatic-hydrothermal origin (Cheng et al. 2024; Figure 9).

At deeper crustal levels, previous studies (e.g., Heinson et al. (2018), Kirkby et al. (2022), Sheng et al. (2024), Kay et al. (2025)) have observed that some mineral deposit types related to convergent margins show a spatial relationship to crustal-scale conductivity anomalies and their vicinities. The northern Delamerian Orogen and the Curnamona Province are host to a range of mineral deposits and prospects, but not all of convergent margin affinity (Figure 9). At mid- to lower crustal levels, orogenic gold deposits are located around the edges of the ENAC and WNAC, whereas the BHC is underlying the world-class Broken Hill zinc lead deposit, and the smaller Pinnacles deposit. Also, several epigenetic copper deposits seem to be proximal to the margins of major conductors (cf. Kay et al. 2025), and two deposits of magmatic-hydrothermal affinity are located along the southern margin of the WC (Figure 9). These observations contribute to the assessment of spatial correlations between electrical conductivity anomalies and mineral deposit types, and will be further investigated in future work.

## 5 Conclusion

We have used new MT data collected in the Curnamona-Delamerian region to image electrical conductivity structures and to inform the understanding of the crustal and lithospheric architecture within the regional geological context.

Our resistivity model confirms crustal-scale conductive features that have been mapped by AusLAMP models, that is, the prominent Curnamona Province Conductor and the two Nackara Arc conductors, and resolves them in greater detail. The new model reveals that the eastern Nackara Arc (ENAC) conductor continues as the Broken Hill Conductor (BHC) into the Curnamona Province. Regional geological considerations suggest that its formation is most likely linked to rifting/extension in the early Cambrian. Although we recognise the east-west trending Wilcannia Conductor could be a possible continuation of the ENAC-BHC zone, integration with recently acquired deep seismic reflection data and evaluation of the geological setting lead us to suggest that they are not genetically linked. We suggest that the Wilcannia Conductor is younger and most likely is related to late Delamerian (~500 Ma) or Siluro-Devonian magmatism. Finally, these conductivity anomalies may represent large-scale trans-crustal structures that control the emplacement of low volume alkaline ultramafic magmas, and show a spatial relationship with certain mineral deposit types, suggesting a possible control on the distribution and formation of metallogenic provinces/belts in the region. Our model enables us to broadly map the arc component of the Delamerian margin, therefore defining the search space for related mineral systems, which will be investigated in future work.

## Declarations

### Data Availability

Processed data and logistic report from the Curnamona Cube Extension survey have been released by Geoscience Australia via https://doi.org/10.26186/147904.

The 3D resistivity model derived from the Curnamona Cube and the Curnamona Cube Extension survey data has been released by Geoscience Australia via https://dx.doi.org/10.26186/148623.

### Competing interests
No potential competing interest was reported by the author(s).

### Funding
This work was supported by Geoscience Australia's Exploring For The Future program and Resourcing Australia's Prosperity initiative program.

**Authors' contributions**

WJ and MD: Conceptualization, Methodology, Investigation, Data curation, Interpretation, Visualization, Writing – original draft. RK: Investigation, Interpretation, Writing– review & editing. MN: Methodology, Visualization. AC and YC: Investigation and Interpretation, Writing – review & editing. AH: Methodology, Data curation, Writing – review & editing.

**Acknowledgements**

The authors acknowledge the traditional owners and landholders within the survey region without whose cooperation these data could not have been collected. The Curnamona Cube Extension survey was part of Geoscience Australia' Exploring for the Future Program. This study builds on the Exploring for the Future program through Geoscience Australia's Resourcing Australia's Prosperity initiative. We greatly appreciate that Prof. Graham Heinson from the University of Adelaide has made the Curnamona Cube survey data and AusLAMP data available for this work and provided valuable feedback on our work. We also thank Gary Egbert, Anna Kelbert and Naser Meqbel for providing the access to the ModEM code. The modelling work was undertaken with the assistance of resources from the National Computational Infrastructure (NCI Australia). The open source MTPy package (Krieger and Peacock 2014, Kirkby et al. 2019) was used to plot phase tensor and to generate ModEM input and output files (https://github.com/MTgeophysics/mtpy).

We acknowledge constructive reviews and comments by the editor and reviewers on an earlier version of this manuscript.

# Figures

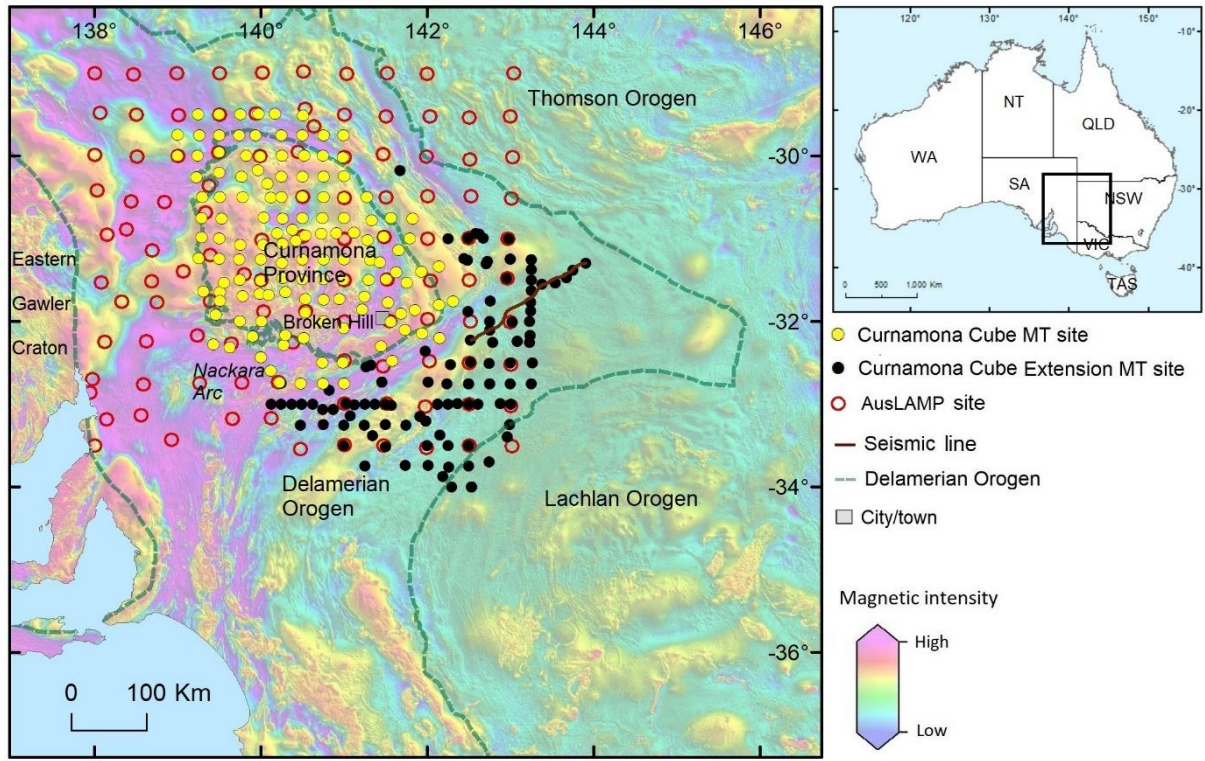

Figure 1 Curnamona Cube and Extension MT survey sites in the Curnamona Province and the adjacent part of the Delamerian Orogen margin. Also shown the new deep-crustal seismic reflection line 22GA-CD2 and AusLAMP sites that are available for this study. The outline of the Delamerian Orogen is based on Raymond (2018). The background map is the reduced to pole magnetic anomaly map of the region (Poudjom Djomani et al. 2019). WA=Western Australia, NT=Northern Territory, QLD=Queensland, SA=South Australia; NSW=New South Wales; VIC=Victoria; TAS=Tasmania.

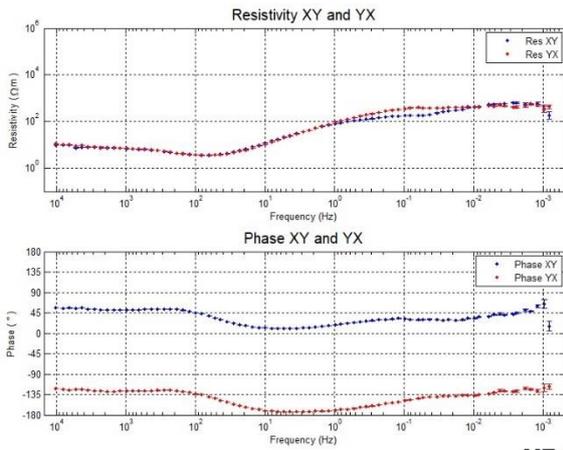
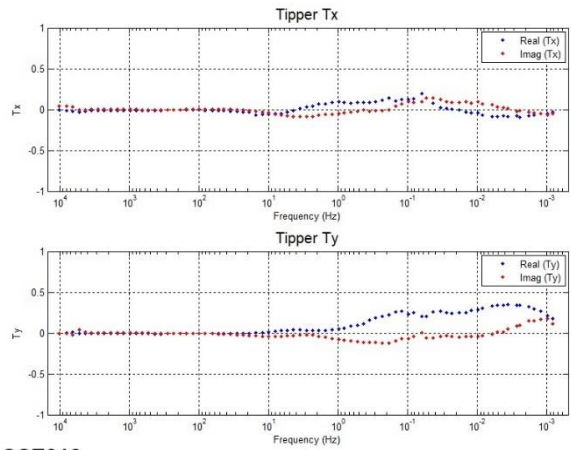
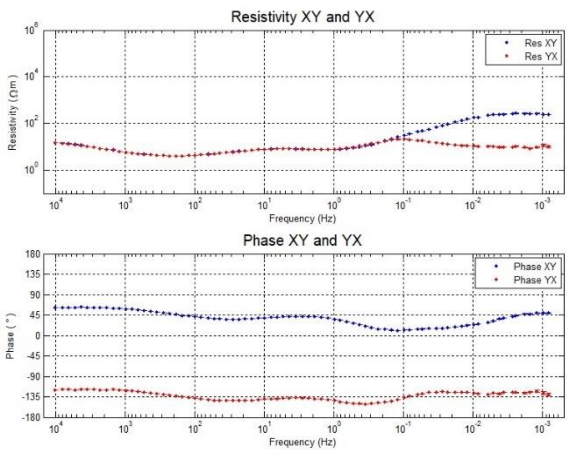
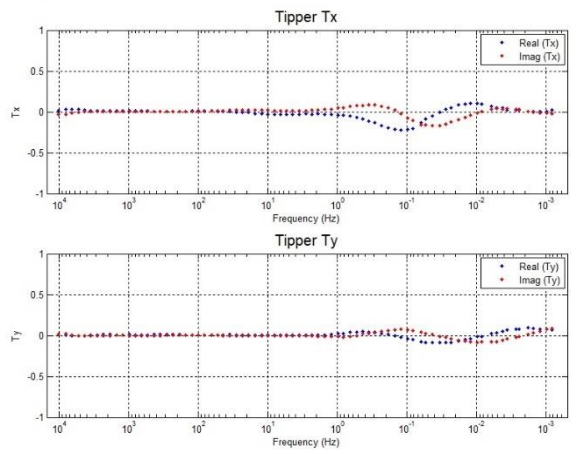
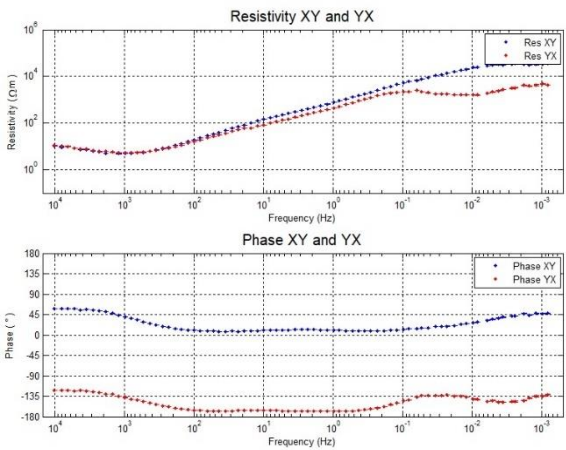
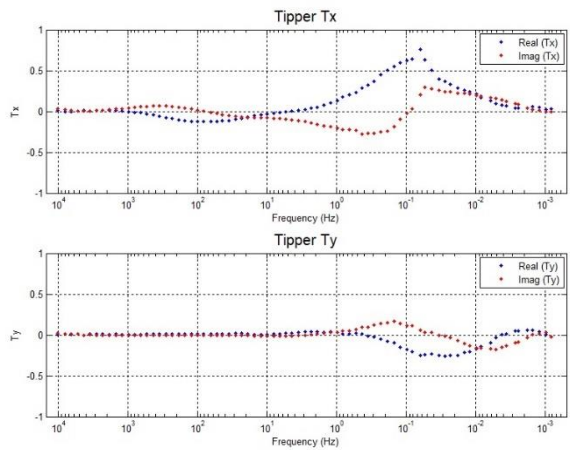

Figure 2 Plots showing processed data (apparent resistivity, phase and vertical transfer function) at site CCE001, CCE016, and CP3 indicating representative data quality and complexity in the survey.

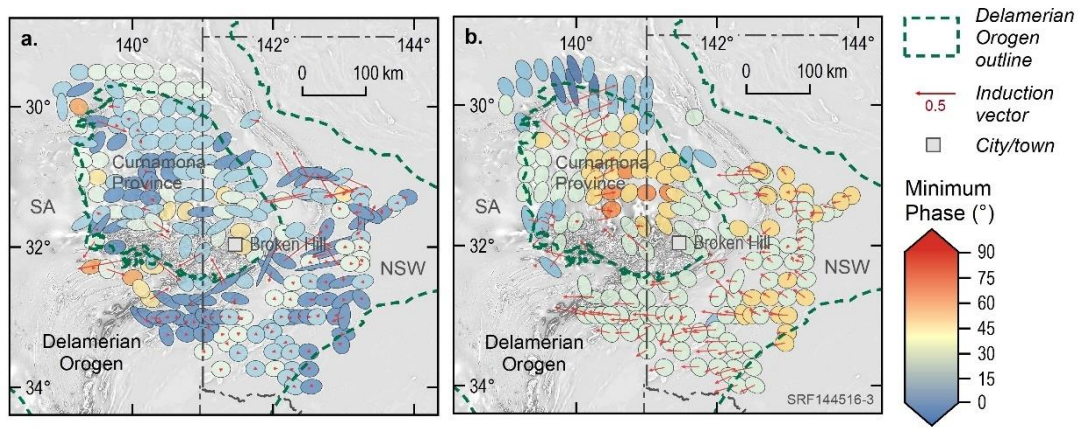

Figure 3 Phase tensor ellipse maps plotted for each site at periods of (a) ~1 s and (b) ~500 s, corresponding to upper and lower crustal depths, approximately. Phase tensor ellipses are coloured by their minimum phase and overlain by induction vectors. The background map is a greyscale image of the reduced to pole magnetic anomaly map of Australia (Poudjom Djomani et al. 2019).

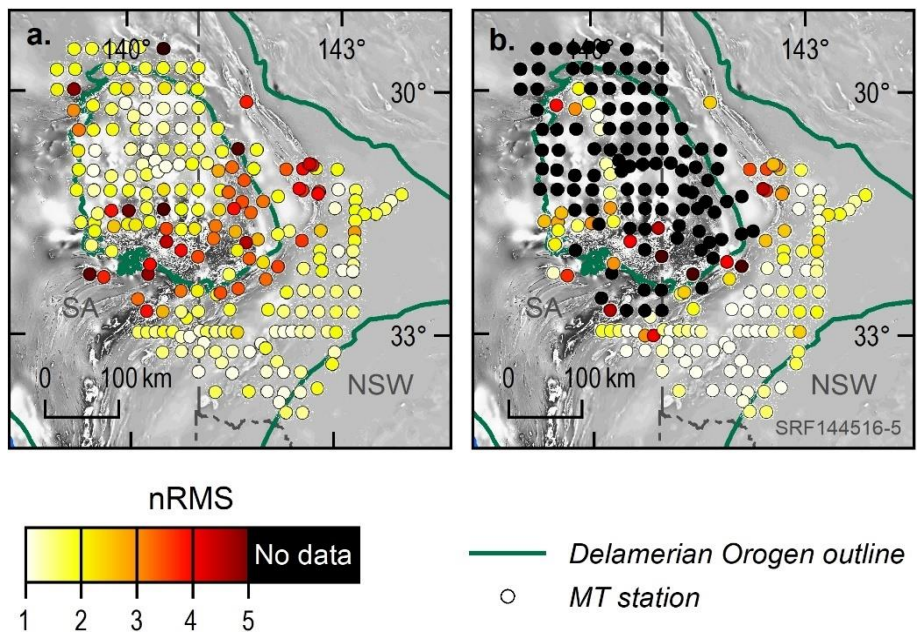

Figure 4 Noise-normalized RMS misfit across all survey sites, plotted for (a) the full impedance tensor and (b) tipper (note, black dots indicate that vertical data are not available at these sites). The background map is a greyscale image of the reduced to pole magnetic anomaly map of Australia (Poudjom Djomani et al. 2019).

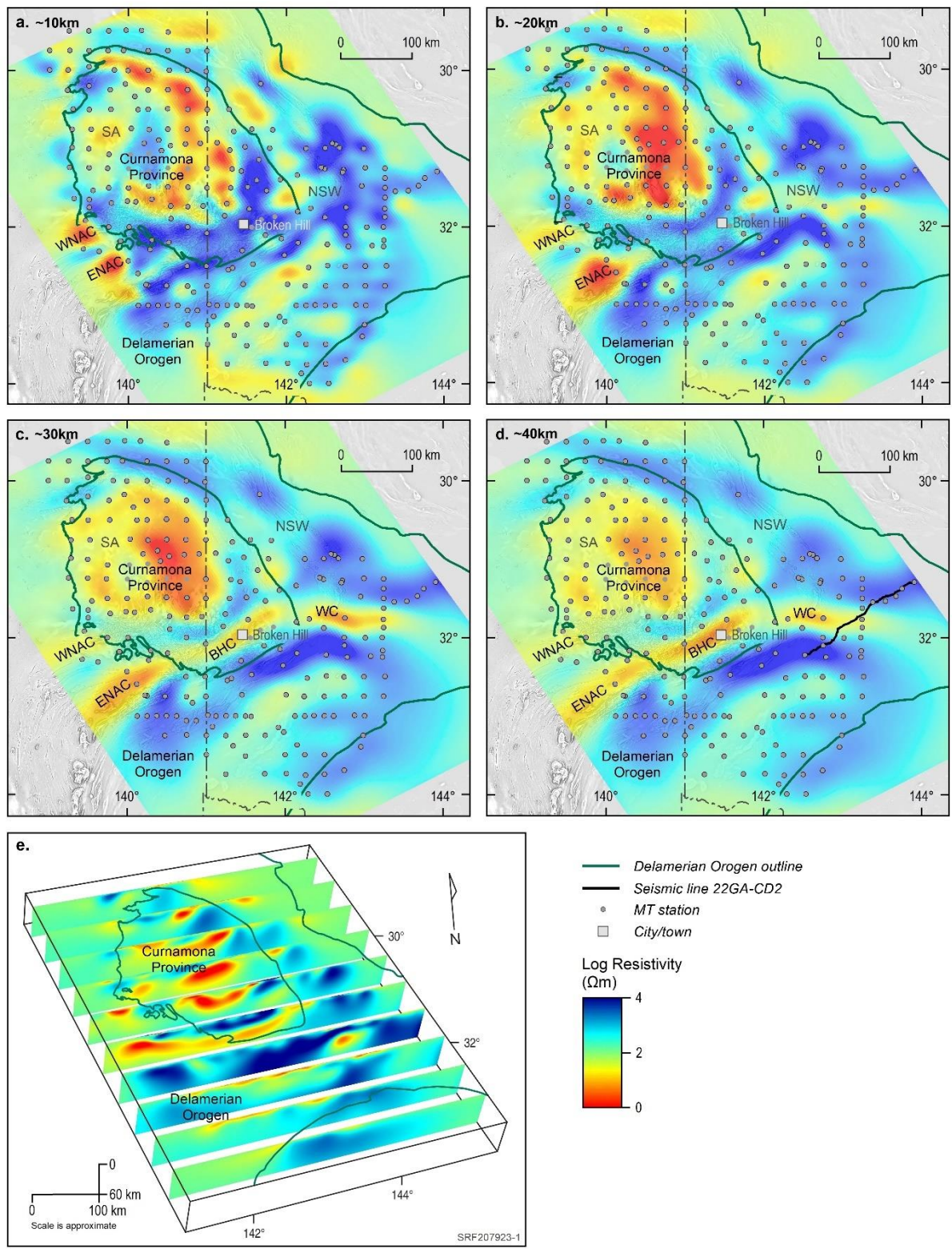

Figure 5 Depth slices from the preferred 3D resistivity model at (a) ~10 km, (b) ~20 km, (c) ~30 km, (d) ~40 km and (e) vertical slices shown within model cartesian space, marked with Delamerian Orogen, Curnamona Province, and Broken Hill. The background map is a greyscale image of the reduced to pole magnetic anomaly map of Australia (Poudjom Djomani et al. 2019). WNAC = Western Nackara Arc Conductor, ENAC = Eastern Nackara Arc Conductor, WC = Wilcannia Conductor, BHC = Broken Hill Conductor.

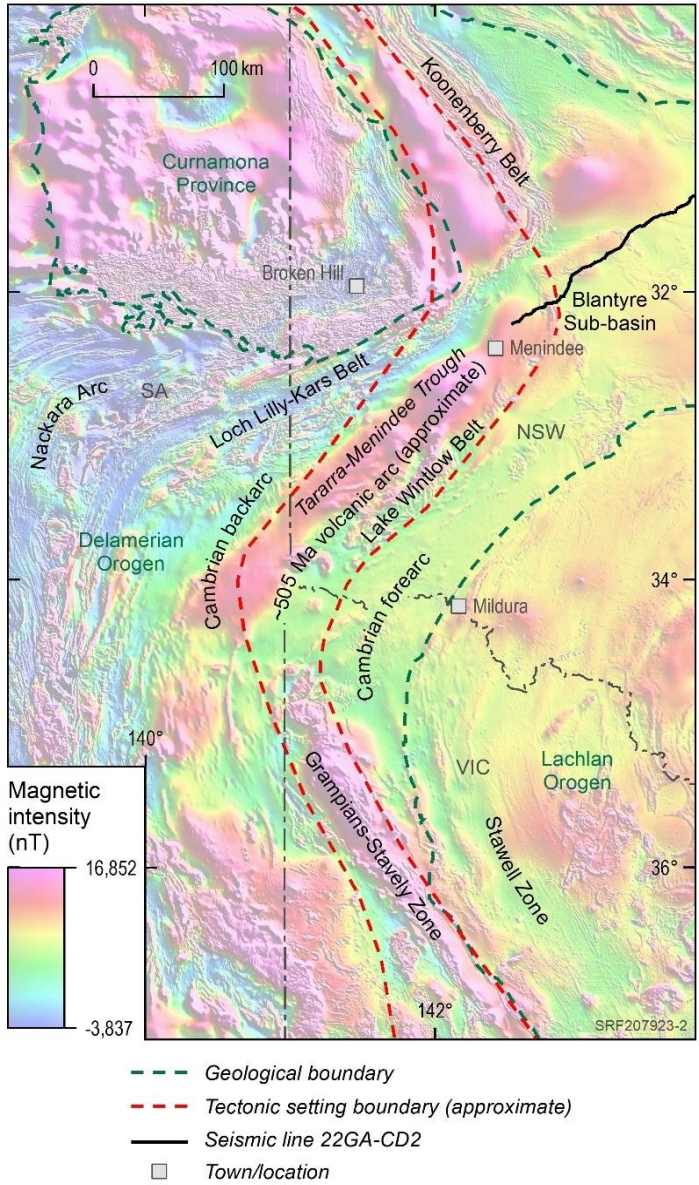

Figure 6 Map of the Curnamona Province and the adjacent part of the Delamerian Orogen. Shown are the orogenic zonation marked in red dashed lines (after Clark et al. 2024), province boundaries, relevant regional geological features, and the position of deep seismic reflection line 22GA-CD2. Background is a colour image from the reduced to pole magnetic anomaly map of Australia (Poudjom Djomani et al. 2019).

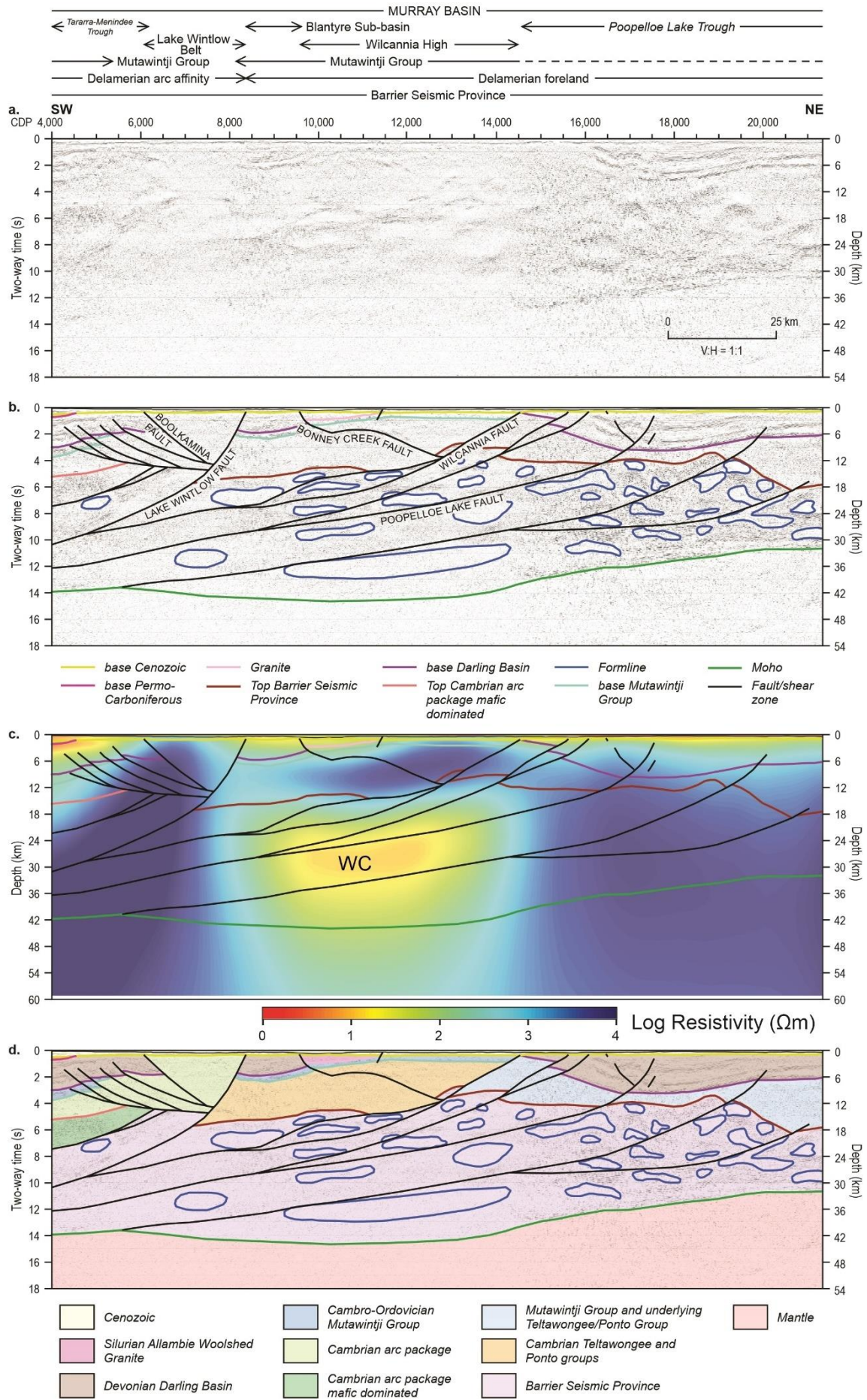

Figure 7 Seismic line 22GA-CD2. (a)Post-Stack Time Migration (poSTM) 18 s TWT display of the uninterpreted data. (b) Interpreted version of the PoSTM deep seismic reflection display shown in (a). (c) 3D resistivity model extracted along seismic line 22GA-CD2 overlain on the interpretation line work. (d) Coloured version of (b) showing main geological units and the Barrier Seismic Province. Vertical to horizontal scale is ~1:1, assuming an average crustal velocity of 6 km/s; CDP – common depth point; WC = Wilcannia Conductor.

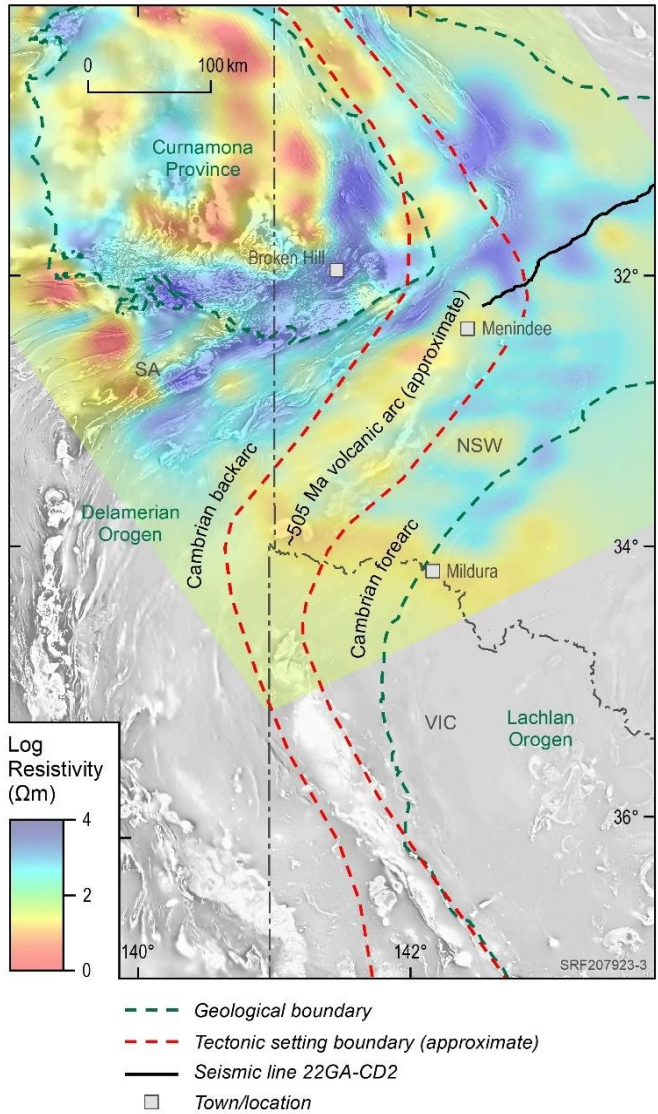

Figure 8 Depth slice at ~10 km from the resistivity model of the Curnamona Province and the adjacent part of the Delamerian Orogen. Shown is the orogenic zonation (after Clark et al. 2024) highlighting the interpreted approximate location of volcanic arc (marked in red dashed lines) in the Delamerian Orogen at ~505 Ma. The background map is a greyscale image of the reduced to pole magnetic anomaly map of Australia (Poudjom Djomani et al. 2019).

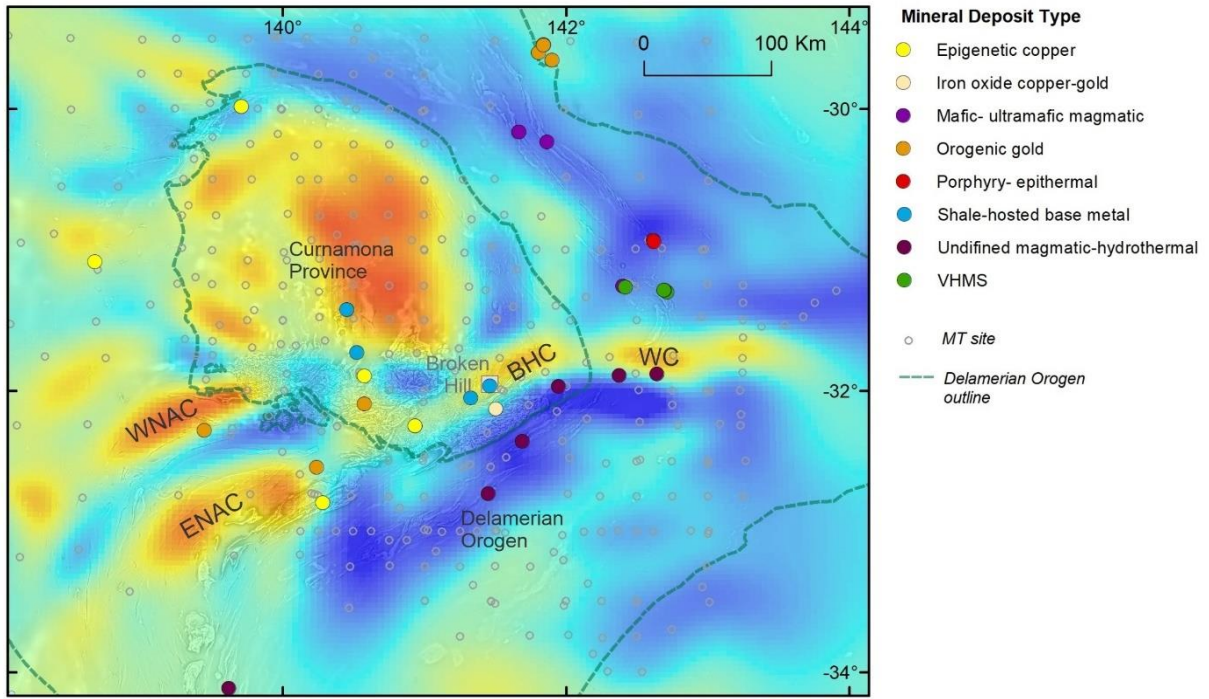

Figure 9 Major mineral deposits overlain on the depth slice of ~30 km from the combined long-period and broadband model presented in the Supplement (Figure S1), marked with Delamerian Orogen, Curnamona Province, and Broken Hill. The background map is a greyscale image of the reduced to pole magnetic anomaly map of Australia (Poudjom Djomani et al. 2019). WNAC = Western Nackara Arc Conductor, ENAC = Eastern Nackara Arc Conductor, WC = Wilcannia Conductor, BHC = Broken Hill Conductor.

# A magnetotelluric image of the Curnamona Province and the adjacent Delamerian Orogen margin – new insights into the crustal architecture

**Supplement**

Wenping Jiang[1], Michael Doublier[1], Russell Korsch[1], Andy Clark[1], Malcolm Nicoll[1], Adrian Hitchman[1], Yanbo Cheng[1]

[1]Geoscience Australia, Canberra, Australian Capital Territory, Australia

Corresponding author: Wenping Jiang (wenping.jiang@ga.gov.au)

**Combined Long-period and Broadband Model**

We have inverted long-period data (10-10,000 s) at 110 AusLAMP sites that are available to this study together with the broadband data to better constrain the crustal and possibly lithospheric scale structures in the region. The site locations are shown in Figure 1 in the main paper. In total, data at 341 sites including impedance tensors and tippers at 35 periods (5/decade) in the range of 0.001–10,000 s was inverted, where data is available. The model setup and parameterisation are similar to the broadband model presented in the main paper, except that the computational mesh was not rotated and was comprised of 166 × 171 horizontal cells of 4 km with 150 vertical layers extending to a much greater depth. For the sites that fall in the same cell, long-period and broadband data was combined into one data file. The starting model was a uniform 500 Ω·m half space. Error floors of 5% of the square root of ZxyZyx (where Zxy and Zyx are the two off-diagonal impedance tensor components) were applied to each impedance tensor component, and an absolute value of 0.02 was applied to the tipper. A covariance value of 0.3 was applied twice in all directions across the model cells and was increased to 0.5 after 109 iterations to further smooth the model. The final model converged to a noise-normalised root-mean-square (RMS) misfit value of 2.56 after a total of 266 iterations. The resistivity model is presented in Figure S1, including depth slices from ~10 km to ~60 km on a 10 km interval.

The combined model confirms the main structures revealed in the broadband model (Figure 5**Error! Reference source not found.**), including the Curnamona Conductor, Nackara Arc conductors, Broken Hill Conductor and Wilcannia Conductor. With more data and better coverage, the combined model resolved the structures with increased constraints, indicating that these features discussed in the main paper are robust.

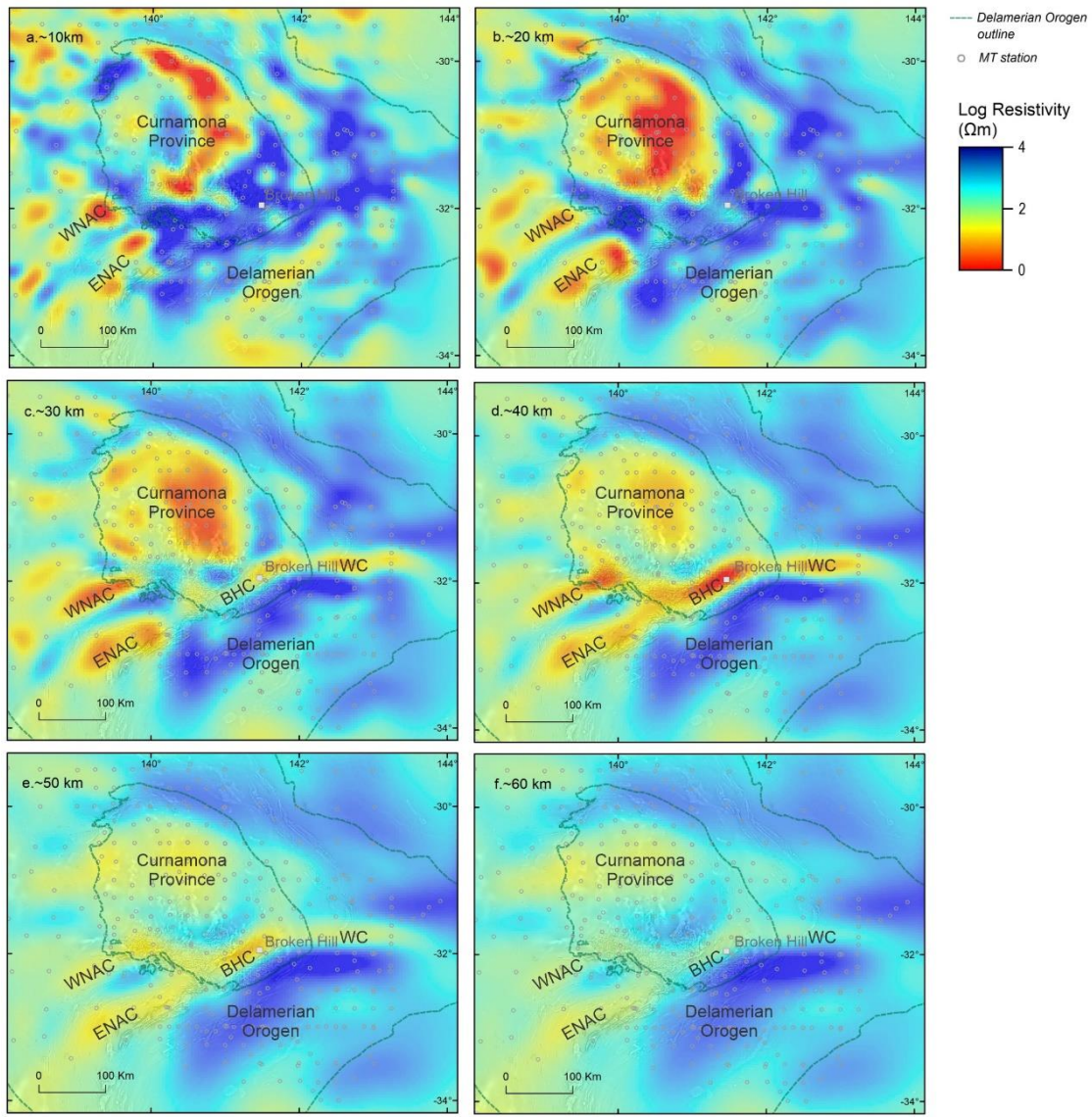

*Figure S1 Depth slices from inverting long-period and broadband data at 341 sites: (a) ~10 km, (b) ~20 km, (c) ~30 km, (d) ~40 km, (e) ~50 km and (f) ~60 km, shown within model cartesian space, marked with Delamerian Orogen, Curnamona Province, and Broken Hill. The outline of the Delamerian Orogen is based on Raymond (2018). The depth slices are draped on a greyscale image of the reduced to pole magnetic anomaly map of Australia (Poudjom Djomani et al. 2019). WNAC = Western Nackara Arc Conductor, ENAC = Eastern Nackara Arc Conductor, BHC = Broken Hill Conductor, WC = Wilcannia Conductor.*

**Model Parameterisation and Sensitivity Tests**

Multiple modelling tests were performed to determine the optimal model parameters and to ensure the robustness of major structures. These include mesh resolution and rotation, starting model, data subset, smoothing factor/covariance, and error floor. In terms of cell size, although a fifth of the site spacing is often used as the rule of thumb (Bedrosian and Feucht 2014, Meqbel et al. 2014, Miensopust 2017), it can be slightly compromised for denser site spacings for regional broadband MT surveys. The principle is to use a suitable resolution to capture the complexity/variation of resistivity structures. In this study, we optimised cell sizes of 3 km for the broadband model and 4 km for the combined model, respectively. This is to accommodate interstation features between denser sites (~10 to 20 km) along the seismic line while maintaining reasonable computational costs. Both models resolved the common structures with negligible difference, shown in Figure 5 **Error! Reference source not found.**and Figure S1. In addition, mesh rotation does not affect the orientations of the major structures, given that data was rotated accordingly.

We inverted the broadband data with three different starting models (uniform half space of 100 Ω·m, 500 Ω·m, and 1000 Ω·m) to test the robustness of major structures. The resultant models recovered the same structures with negligible difference, although the number of iterations taken by the models varied. The inversion model starting from 1000 Ω·m half space took the greatest number of iterations and achieved the best model/data fit (RMS). This demonstrates that inversion is ultimately driven by data. Smoothing factor is often utilised as a trade-off between model roughness and data/model fit, and geologically, there is little difference in the major structures derived from the models, as expected on a regional scale.

We also inverted the full impedance tensor only. As shown in Figure S2, the model recovered the prominent features, however, their lateral boundaries are not well defined due to lack of constraints from the tipper data. At lower crust (~30 km), the southwestern part of Broken Hill Conductor diverged from the northeast to east-northeast trend, extending into the Curnamona conductor following a curvilinear path. Most likely it is an artifact created by a few localised MT sites without lateral constraints from the tipper data. This highlights the importance of a joint inversion of full impedance tensor and tipper, and their respective contributions.

A more comprehensive sensitivity analysis of 3D MT inversion model setup, parameters, and data subsets using ModEM can be found in Robertson et al. (2020).

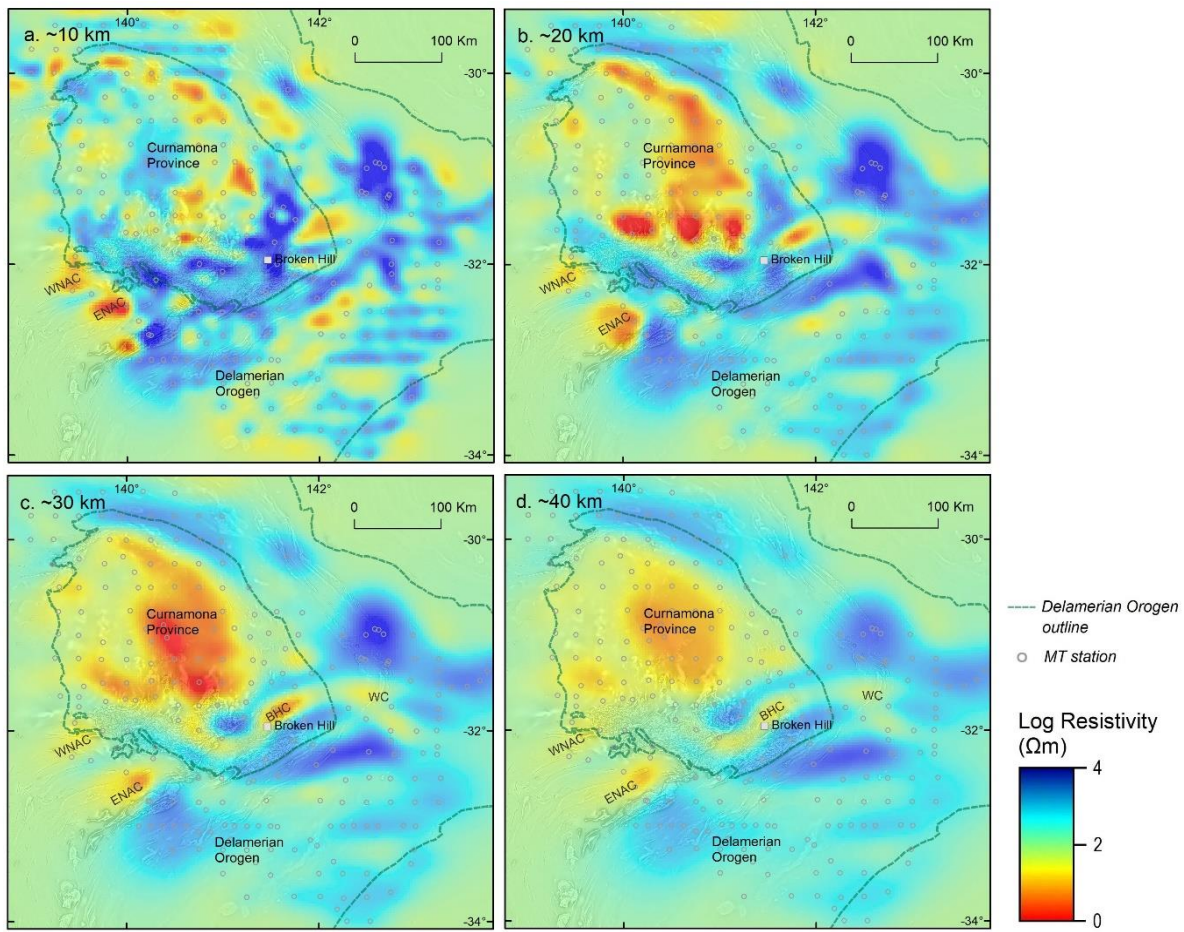

*Figure S2 Depth slices from inverting the impedance tensor only: (a) ~10 km, (b) ~20 km, (c) ~30 km and (d) ~40 km, shown within model cartesian space, marked with Delamerian Orogen, Curnamona Province, and Broken Hill. The outline of the Delamerian Orogen is based on Raymond (2018). The depth slices are draped on a greyscale image of the reduced to pole magnetic anomaly map of Australia (Poudjom Djomani et al. 2019). WNAC = Western Nackara Arc Conductor, ENAC = Eastern Nackara Arc Conductor, BHC = Broken Hill Conductor, WC = Wilcannia Conductor.*

**Model Response vs. Data**

For the preferred model, the RMS misfit plot (Figure 4) shows a reasonably good fit across most of the area, except the southern edge of the Curnamona Province. In part, this is due to the larger site spacing and lower model resolution, compared to the higher complexity of geology and variation of the resistivity structures. Figure S3 gives comparisons of observed data and model responses from the preferred model at nine representative sites. The site location coordinates can be found in Jiang et al. (2023). In general, the apparent resistivity off-diagonal components ($\rho_{xy}$ and $\rho_{yx}$) fit better than the diagonal components ($\rho_{xx}$ and $\rho_{yy}$) because of the higher magnitude in high frequencies. Sites CCE003, Cube_127 and Cube_132 are located at the southern edge of the Curnamona Province. At these sites, both the diagonal components of apparent resistivity and phase fit poorly. Note, vertical data was not collected at site Cube_122.

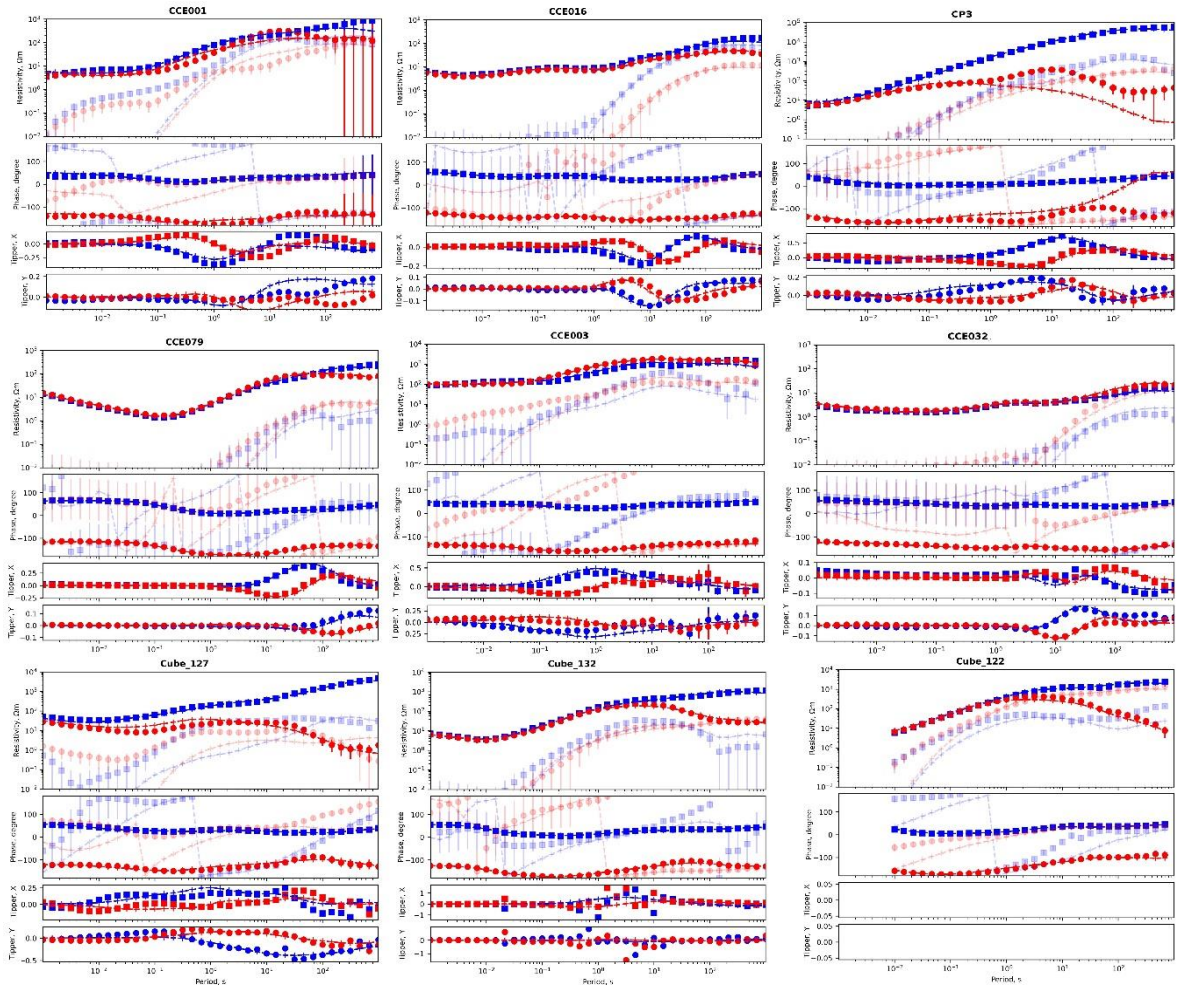

Figure S3 Comparisons of data (squares/points) and model responses (dashed lines) from the preferred inversion model at nine representative sites, including apparent resistivity, phase, and vertical transfer functions (tipper). The $\rho_{xx}$ and $\rho_{xy}$ components of apparent resistivity and Real components of the tipper are shown in blue, the $\rho_{yx}$ and $\rho_{yy}$ components of and Imaginary components are shown in red. The diagonal components of apparent resistivity and phase are semi-transparent.